\documentclass[12pt]{article}
\usepackage{amssymb}
\usepackage{epsfig,color}
\usepackage{pstricks,graphicx,epsfig,color,amssymb,amsmath,amscd}
\usepackage{cite}
\usepackage[]{graphicx}
\usepackage{makeidx}
\usepackage{amsmath}
\usepackage{nccmath}
\usepackage{setspace}
\usepackage{ulem} 
\usepackage{tensor}
\usepackage{bm}

\usepackage[format=hang, justification=justified, font=small, labelsep=period
]{caption} 
\renewcommand\rho{\varrho}

\newcommand{\be}{\begin{eqnarray}}
\newcommand{\ee}{\end{eqnarray}}

\newcommand{\lambdabar}{{\mkern0.75mu\mathchar '26\mkern -9.75mu\lambda}}

\numberwithin{equation}{section}

\begin{document}
\begin{titlepage}
\title{Conversion of protons to positrons by a black hole}
\author{A.\,D. Dolgov$^{a, b,}$\footnote{dolgov@nsu.ru} \ and A.\,S. Rudenko$^{a, c,}$\footnote{a.s.rudenko@inp.nsk.su}}
\date{}

\maketitle
\begin{center}
$^a${Department of Physics, Novosibirsk State University, \\ 
Pirogova st.\,2, Novosibirsk, 630090 Russia} \\
$^b${Bogoliubov Laboratory of Theoretical Physics, Joint Institute for Nuclear Research, \\
Joliot-Curie st.\,6, Dubna, Moscow oblast, 141980 Russia} \\
$^c${Theory Division, Budker Institute of Nuclear Physics, \\
akademika Lavrentieva prospect 11, Novosibirsk, 630090 Russia} 
\end{center}

\begin{abstract}

The conversion of protons to positrons at the horizon of a black hole (BH) is considered. It is shown that the process may efficiently proceed for BHs with masses in the range $\sim 10^{18}$\,---\,$10^{21}$ g. It is argued that the electric charge of BH acquired by the proton accretion to BH could create electric field near BH horizon close to the critical Schwinger one. It leads to efficient electron-positron pair production, when electrons are back captured by the BH while positrons are emitted into outer space. Annihilation of these positrons with electrons in the interstellar medium may at least partially explain the origin of the observed 511 keV line.

\end{abstract}
\thispagestyle{empty}
\end{titlepage}

\section{Introduction} \label{intro}

As has been noticed many years ago \cite{Shv} (see also \cite{TZTI, ZTT} and references therein) the accretion of ionized interstellar matter onto a neutron star leads to positive electric charging of the star. The reason is that the pressure of the emerging radiation acts efficiently only on electrons\footnote{Thomson cross section for protons is $(m_p/m_e)^2 \approx 3 \times 10^6$ times smaller than for electrons.}, while proton motion is determined mostly by gravity. Therefore, the generated electric field equalizes the flows of electrons and protons falling onto a neutron star.

If one considers a nonrotating black hole (BH) instead of a neutron star, the above conclusions do not change \cite{TT, BDP}. However, due to the smallness of gravitational radius $r_g$ for not very massive BH the electric field at the horizon of BH can be strong enough, $E \sim Q_{BH}/r_g^2$, and the $e^+ e^-$ pairs can be produced by Schwinger mechanism \cite{Schw} near the BH horizon. Then electrons are back-captured due to strong Coulomb attraction in the vicinity of the BH and thus the flux of the electrons from BH can be neglected, while positrons are rejected away by the Coulomb repulsion. 

Existence of rich populations of positrons in the Galaxy was noticed long ago through the observations of 511 keV gamma ray line (see~\cite{anti-e1, anti-e2, anti-e3} and references therein) with the flux from the Galactic bulge
\be
\Phi_\text{511 keV}^\text{bulge} \sim 10^{-3} \text{ photons/(cm$^2 \, \cdot$ s)}, 
\ee
that unambiguously indicates the frequent annihilation of nonrelativistic $e^+ e^-$ pairs in the Galactic bulge with the rate~\cite{anti-e1}
\be \label{Neebulge}
\dot N_{ee}^\text{bulge} \sim 10^{43} \text{ s}^{-1}.
\ee
Note that one of the brightest X-ray point sources in the region around the Galactic Center, where from 511 keV photons are coming, got the name Great Annihilator~\cite{gr-ann}. Possibly it is a microquasar first detected in soft X-rays by the Einstein Observatory~\cite{Ein-observ} and later detected in hard X-rays by the space observatory ``Granat''~\cite{granat}.

There is no commonly accepted point of view on the origin of the cosmic positrons. The conventional hypothesis that positrons are created in strong magnetic fields of pulsars is at odds with the AMS data~\cite{AMS-19}. One more possibility is that positrons are primordial, produced in the early universe in relatively small antimatter domains~\cite{DS, DKK, BBBDP}. Possible observation of the unexpectedly high flux of antinuclei~\cite{anti-nuc-AMS-1, anti-nuc-AMS-2} and antistars in the Galaxy~\cite{anti-stars} strongly supports this hypothesis. As argued in \cite{bpbbd}, antihelium cosmic rays are created by antistars.

In this paper we study the described above alternative possibility that positrons are created by the Schwinger process at the BH horizon. This mechanism was described in \cite{BDP}, but here we modify, correct, and made more detailed analysis of the previous calculations.

We consider here BHs with masses above $10^{17}$ g, since BHs with smaller masses and density equal to the density of dark matter would create too large flux of positrons through the Hawking evaporation~\cite{Laha-1, Laha-2, Carr-Kuhnel-1, Carr-Kuhnel-2}. BHs with masses $M > 10^{17}$ g do not create sufficient flux of positrons to explain the observed flux of 511 keV photons.

We assume that the BH masses are close to $10^{20}$ g within a couple of orders of magnitude, and parametrise the mass as 
\be
M = M_{20} \times 10^{20} \text{ g}.
\ee
This choice is dictated by the conditions that for such BHs the generated electric field at horizon would be close to the critical Schwinger value \cite{TT} and that such primordial BHs might make significant contribution to the cosmological mass density, even making 100\,\% of dark matter~\cite{Carr-Kuhnel-1, Carr-Kuhnel-2}. One more condition is that the electric field of such BHs should be sufficiently small, so that the electric repulsion of protons is weaker than their gravitational attraction. Denote the radius of the corresponding stellar object as $R$ (at the moment we do not say that it is a black hole). The electric field on the surface $r = R$ should be close to the critical Schwinger one:
\be
E(R) = \frac{\alpha Q}{R^2} \sim E_c = \frac{m_e^2}{\sqrt {4 \pi \alpha}},
\label{E-of-R}
\ee
where $Q$ is the electric charge of BH in units of the proton charge $e = \sqrt \alpha$.

The force of the gravitational attraction would be stronger than that of the Coulomb repulsion if
\be
\alpha Q < \frac{M m_p}{m_{Pl}^2},
\label{el-over-grav}
\ee
where $m_{Pl}$ is the Planck mass and $m_p$ is the proton mass.
 
Conditions \eqref{E-of-R} and \eqref{el-over-grav} can be rewritten as
\be
M \lesssim \frac{\sqrt \alpha m_{Pl}^2 m_p}{m_e^2} \left( \frac{r_g}{R} \right)^2 \sim 10^{20} \left( \frac{r_g}{R} \right)^2 \text{ g}.
\label{M-smaller}
\ee
Let us note that this relation is approximate and the electric field may be noticeably smaller than the critical Schwinger one \eqref{E-of-R} but still efficiently create electron-positron pairs due to huge pre-exponential factor.

In our calculations we use natural units $\hbar = c = k_B = 1$ and the following constants: \\
fine-structure constant: $\alpha = e^2 \approx 1/137$, \\
electron mass: $m_e \approx 0.5$ MeV, \\
proton mass: $m_p \approx 938$ MeV, \\
Planck mass: $m_{Pl} \approx 1.2 \times 10^{22}$ MeV, \\
gravitational radius of BH with mass $M$: $r_g = 2 M/m_{Pl}^2 \approx 1.5 \times 10^{-8} M_{20}$ cm.

\section{Accretion onto a black hole} \label{accretion}

We consider spherical accretion of ionized hydrogen, i.\,e. protons and electrons, onto a nonrotating BH. In a region not too close to the BH, $r \gg r_g \sim 10^{-8}$ cm, we can assume that the particles are nonrelativistic. As usual, we choose the origin of coordinate system at the center of the BH and direct the axis $\bm r$ along the BH radius outward. Equations of motion in such case look as follows \cite{BDP}: \\
\begin{align}
m_p \dot v_p & = - \frac{G_N M m_p}{r^2} + \frac{\alpha Q}{r^2} + \frac{L \sigma_{\gamma p}}{4 \pi r^2} - \sigma_{\gamma p} n_\gamma \omega_\gamma v_p - \sigma_{e p} n_e P_p (v_p - v_e), 
\label{dotvp-corr} \\
m_e \dot v_e & = - \frac{G_N M m_e}{r^2} - \frac{\alpha Q}{r^2} + \frac{L \sigma_{\gamma e}}{4 \pi r^2} - \sigma_{\gamma e} n_\gamma \omega_\gamma v_e - \sigma_{e p} n_p P_e (v_e - v_p).
\label{dotve-corr}
\end{align} 
Here the first and second terms on the right hand side of the equations are gravitational and Coulomb forces, respectively, where $G_N = 1/m_{Pl}^2$ is the gravitational constant and $Q$ is the electric charge of BH in units of the proton charge $e = \sqrt \alpha$. The third terms describe the interactions of particles with the photons emitted by a BH, where $L = \eta \dot M$ is accretion luminosity with $\eta$ being radiative efficiency of accretion, $\sigma_{\gamma p} = 8 \pi \alpha^2/(3 m_p^2)$ and $\sigma_{\gamma e} = 8 \pi \alpha^2/(3 m_e^2)$ are the Thomson cross sections. The fourth terms correspond to the interactions of particles with photons in cosmic plasma, where $n_\gamma$ and $\omega_\gamma$ are the photon number density and the photon energy, respectively. And finally, the fifth terms describe the interaction of protons and electrons with each other, where $\sigma_{e p}$ is cross section of $e p\mspace{1 mu}$-scattering, $n_p$ and $n_e$ are the proton and electron number densities, $P_p$ and $P_e$ are the average values of proton and electron momenta, respectively.

Since the mean free path of electrons in cosmic plasma is very large, $\lambda_e \sim 1/(\sigma_{\gamma e} n_\gamma) \sim 5 \times 10^{10}\ (1 \text{ eV}/T_\gamma)^3$ cm (where temperature of photon gas $T_\gamma$ is usually $\lesssim 1$ eV)\footnote{Here $n_\gamma = 2 \zeta (3) T_\gamma^3/\pi^2 \approx 0.24\, T^3_\gamma$.}, and for protons the mean free path is even much larger, $\lambda_p = (m_p/m_e)^2 \lambda_e \gg \lambda_e$, we can neglect the fourth terms in \eqref{dotvp-corr} and \eqref{dotve-corr} for $r \ll \lambda_e$. Due to the fact that the fifth terms in \eqref{dotvp-corr} and \eqref{dotve-corr} include the difference between the velocities of protons and electrons, these terms are also small. As we will find below, the relative velocity difference $(v_p - v_e)/v_e$ is less than $10^{- 5}$, e.\,g. see Fig.~\ref{fig_epsM20}. That is, protons and electrons fall into the BH at approximately the same velocities. So assuming that factors $\sigma_{e p} n_{e, p} P_{p, e}$ are not very large at the distances under consideration, the fifth terms in \eqref{dotvp-corr} and \eqref{dotve-corr} could be neglected.

Note that we do not take into account the equation for positrons as well as corresponding proton-positron and electron-positron interaction terms in \eqref{dotvp-corr} and \eqref{dotve-corr} because positron density is negligible at the distances under consideration.

Let us also make a remark explaining why we consider Eqs. \eqref{dotvp-corr}, \eqref{dotve-corr} and do not use the classical Bondi result \cite{Bondi} for the accretion rate. This is because Bondi model is not quite relevant to the problem at hand. Indeed, we consider charged black holes, whose charge is formed due to the fact that the pressure of the emerging radiation acts efficiently only on electrons, while proton motion is determined mostly by gravity. Therefore, we should consider different equations of motion for protons and electrons. Bondi model does not take into account this effect of charge separation and treats accreting substance as neutral gas. And the second reason is that we consider equations of motion for protons and electrons at the distances $r \ll \lambda_e$. On the contrary, Bondi model is valid for distances $r \gg \lambda_e$ and treats the gas as a continuous fluid, having velocity, temperature and density defined at each point. Accretion radius in Bondi model is
\[
r_\text{acc} = \frac{2 G_N M}{c_\text{s}^2 (\infty)} \simeq 3 \times 10^{14} \left( \frac{M}{M_\odot} \right) \left( \frac{10 \text{ km/s}}{c_\text{s} (\infty)} \right)^2 \text{ cm},
\]
where $c_\text{s} (\infty)$ is the speed of sound in the gas at ``infinity'' (in a warm region it is on the order of 10 km/s) and $M_\odot \simeq 2 \times 10^{33}$ g is the solar mass. Therefore, for the black holes of solar masses one has $r_\text{acc} \gg \lambda_e$ and Bondi model is valid, but for the black holes under consideration ($M \sim 10^{20}$ g) one has $r_\text{acc} \sim 10 \text{ cm } \ll \lambda_e$ and Bondi model is not applicable.

After all the comments made, let us move on to solving the system of Eqs. \eqref{dotvp-corr}, \eqref{dotve-corr}. In the region $r \gg r_g$,  where both protons and electrons are nonrelativistic and remembering that $r \ll \lambda_e$, we can reduce this system to the following one:
\begin{align}
m_p \dot v_p & \approx - \frac{G_N M m_p}{r^2} + \frac{\alpha Q}{r^2} + \frac{L \sigma_{\gamma p}}{4 \pi r^2} = - \frac{\beta_p}{r^2}, 
\label{dotvp2} \\
m_e \dot v_e & \approx - \frac{G_N M m_e}{r^2} - \frac{\alpha Q}{r^2} + \frac{L \sigma_{\gamma e}}{4 \pi r^2} = - \frac{\beta_e}{r^2}, 
\label{dotve2}
\end{align}
where 
\be
\beta_p & = & G_N M m_p - \alpha Q - L \sigma_{\gamma p}/(4 \pi) \approx G_N M m_p - \alpha Q, 
\label{betap} \\
\beta_e & = & G_N M m_e + \alpha Q - L \sigma_{\gamma e}/(4 \pi). 
\label{betae}
\ee

Equations \eqref{dotvp2} and \eqref{dotve2} can be easily integrated
\begin{align}
\frac{m_p v_p^2}{2} & \approx \frac{\beta_p}{r} + C_p, \label{Ep} \\
\frac{m_e v_e^2}{2} & \approx \frac{\beta_e}{r} + C_e, \label{Ee}
\end{align}
where $C_p$ and $C_e$ are integration constants.

Since we consider the case when protons fall into a BH, $m_p \dot v_p < 0$ and consequently $\beta_p > 0$, then we have
\be \label{condition_p}
\alpha Q < G_N M m_p - \frac{L \sigma_{\gamma p}}{4 \pi} \approx G_N M m_p = \frac{M m_p}{m_{Pl}^2} \approx 3.5 \times 10^5\, M_{20}.
\ee
So, it is convenient to parametrise the BH charge $Q$ as 
\be \label{q}
q = \frac{\alpha Q}{m_p r_g} = \frac{\alpha Q m_{Pl}^2}{2 m_p M} \approx 10^{-8}\, \frac{Q}{M_{20}}.
\ee
Then \eqref{condition_p} leads to
\be \label{condition_q_p}
q < q_{\max} = \frac{1}{2}.
\ee
Otherwise protons would fly away from the BH.

The condition that electrons also fall into the BH is as follows
\be \label{condition_e}
\beta_e > 0 \quad \Rightarrow \quad \alpha Q > \frac{L \sigma_{\gamma e}}{4 \pi} - G_N M m_e \quad \Rightarrow \quad q > \frac{\alpha^2 \eta}{3}\, \frac{m_{Pl}^2}{m_e^2}\, \frac{\dot N_p}{M} - \frac{m_e}{2 m_p},
\ee
where we take into account that accretion luminosity $L = \eta \dot M \approx \eta m_p \dot N_p$, because increase of BH mass occurs mainly due to protons.

Since the produced positrons fly away from the BH, then $\beta_{e^+} < 0$, where $\beta_{e^+} = G_N M m_e - \alpha Q - L \sigma_{\gamma e}/(4 \pi)$, and consequently
\be \label{condition_pos}
q > \frac{m_e}{2 m_p} - \frac{\alpha^2 \eta}{3}\, \frac{m_{Pl}^2}{m_e^2}\, \frac{\dot N_p}{M}.
\ee
Combining the inequalities \eqref{condition_q_p}\,---\,\eqref{condition_pos}, we obtain the following constraint on the BH charge:
\be \label{condition_final}
\left|\, \frac{\alpha^2 \eta}{3}\, \frac{m_{Pl}^2}{m_e^2}\, \frac{\dot N_p}{M} - \frac{m_e}{2 m_p}\, \right| < q < \frac{1}{2}.
\ee

We assume that the process of BH charging tends to the stationary one, i.\,e. the BH charge does not change asymptotically, or in other words, the total numbers of protons and electrons\footnote{Including both electrons falling into a BH from the cosmic plasma from relatively large distances, and electrons produced by the Schwinger mechanism near the BH.} falling into a BH per unit time are equal. Indeed the electric field of BH creates $e^+ e^-$ pairs, but the positrons are rejected away by the Coulomb repulsion, while the produced electrons are back-captured by the BH due to strong Coulomb attraction and gravity. Therefore, the number of electrons from the cosmic plasma that fall into the BH from the region $r \gg r_g$ is less than the number of protons, so we can introduce the corresponding proton-to-electron ratio
\be \label{k}
k = \dot N_p/\dot N_e \geq 1,
\ee
where $\dot N_p$ and $\dot N_e$ are the fluxes\footnote{The number of particles passing through a given surface per unit time.} of protons and cosmic plasma electrons (not the electrons produced by BH\,!), respectively, captured by the BH:
\be \label{dotNpe}
\dot N_{p, e} = 4 \pi r_g^2 n_{p, e}(r_g) v_{p, e}(r_g) \approx 4 \pi r^2 n_{p, e}(r) v_{p, e}(r).
\ee
Here we suppose that during spherical accretion the fluxes are approximately constant, i.\,e. we use the continuity equations, $\dot N_{p, e} = \mathrm{const}$. In other words, taking into account the equations of motion of protons and electrons at distances far away from the BH\footnote{Where we can consider that the particles are nonrelativistic and particle number densities are not large, as well as can neglect the fourth and fifth terms in \eqref{dotvp-corr} and \eqref{dotve-corr}.} we calculate the fluxes at such distances. Then we assume that due to gravitational attraction of the BH, all these protons and electrons fall into the BH. Thus, due to the flux conservation we can consider the calculated fluxes as fluxes of protons and electrons captured by the BH.

Let us assume that far from BH ($r \gg r_g$) cosmic plasma is electrically neutral, so number densities of protons and electrons are equal, $n_p(r) = n_e(r)$, then from \eqref{k}, \eqref{dotNpe} we have that $v_p (r) = k v_e (r)$, and from \eqref{Ep}, \eqref{Ee} that 
\[
\frac{\beta_p}{m_p} \approx k^2 \frac{\beta_e}{m_e}.
\]
Here we neglect the integration constants $C_{p, e} \approx m_{p, e} v_{p, e}^2(\lambda_{p, e})/2 - \beta_{p, e}/\lambda_{p, e}$ which are small compared to $\beta_{p, e}/r$ for $r \ll \lambda_{p, e}$. Consequently, from \eqref{betap} and \eqref{betae} we get
\begin{align}
& \alpha Q \bigg(\, \underset{\text{small}}{\underbrace{\frac{1}{m_p}}} + \frac{k^2}{m_e} \bigg) = - \left( k^2 - 1 \right) G_N M + \frac{L}{4 \pi} \bigg( \frac{k^2 \sigma_{\gamma e}}{m_e} - \underset{\text{small}}{\underbrace{\frac{\sigma_{\gamma p}}{m_p}}}\, \bigg) \quad \Rightarrow \\
\Rightarrow \quad & \alpha Q \approx - \frac{k^2 - 1}{k^2}\, G_N M m_e + \frac{L \sigma_{\gamma e}}{4 \pi} = - \frac{k^2 - 1}{k^2}\, \frac{M m_e}{m_{Pl}^2} + \frac{2 \alpha^2 \eta \dot M}{3 m_e^2}.
\end{align}
Then
\be \label{q_equil}
q = \frac{\alpha Q}{m_p r_g} \approx - \frac{k^2 - 1}{k^2}\, \frac{m_e}{2 m_p} + \frac{\alpha^2 \eta}{3}\, \frac{m_{Pl}^2}{m_e^2}\, \frac{\dot M}{m_p M} < \frac{\alpha^2 \eta}{3}\, \frac{m_{Pl}^2}{m_e^2}\, \frac{\dot M}{m_p M},
\ee
since $k \geq 1$.

Taking into account that $\dot M \approx m_p \dot N_p$ we obtain the following expression for the flux of protons captured by BH:
\be \label{dotNp}
\dot N_p & \approx & \frac{\dot M}{m_p} \approx \frac{3 M}{\alpha^2 \eta}\, \frac{m_e^2}{m_{Pl}^2}\, \left( q + \frac{k^2 - 1}{k^2}\, \frac{m_e}{2 m_p} \right) \approx \nonumber \\
& \approx & \frac{0.1}{\eta}\, M_{20} \left( q + 2.7 \times 10^{-4} \, \frac{k^2 - 1}{k^2} \right) \times 8.4 \times 10^{28} \text{ s}^{- 1}.
\ee
Here we use that a reasonable estimate for radiative efficiency of accretion is $\eta \lesssim 0.1$ \cite{FKR}.

In the next Section we calculate the number of $e^+ e^-$ pairs produced by charged BH per unit time, $\dot N_{ee}$.

\section{Production of $e^+ e^-$ pairs by a charged black hole} \label{s-positron-flux}

Since the gravitational radius of the considered BH is $\sim 10^{-8} M_{20}$ cm, i.\,e. is much larger than the Compton wavelength of electron, $\lambdabar_e = 1/m_e \approx 4 \times 10^{-11}$ cm, we can conclude that the electric field is essentially homogeneous at this scale and assume that the $e^+ e^-$ production proceeds according to the Schwinger result~\cite{Schw}. Therefore, the $e^+ e^-$ pair production probability per unit time and unit volume is 
\be
W (x) = \frac{m_e^4}{4 \pi^3 x^2} \sum_\text{$n = 1$}^\infty \frac{\exp(- \pi n x)}{n^2},
\label{W-of-E}
\ee
where $x (r) = E_c/E (r)$ with 
\be
E_c = \frac{m_e^2}{\sqrt{4 \pi \alpha}}
\label{E-c}
\ee
being critical electric field and $E (r) = Q_{BH}/r^2 = \sqrt \alpha\, Q /r^2$ being the electric field of BH at distance $r$. Here $Q$ is electric charge of BH in units of proton charge $e = \sqrt \alpha$ and is to be calculated in what follows.

Thus the ratio of the critical Schwinger field $E_c$ to the value of the electric field at the BH horizon $E (r_g)$ can be presented as 
\be \label{E-c-to-E-of-rg}
x (r_g) = \frac{E_c}{E (r_g)} = \frac{m_e^2}{\sqrt{4 \pi \alpha}}\, \frac{r_g^2}{\sqrt \alpha\, Q} = \frac{m_e^2 r_g}{\sqrt{4 \pi\,} q m_p} = \frac{m_e^2 M}{\sqrt{\pi\,} q m_p m_{Pl}^2} \approx \frac{0.06 M_{20}}{q}.
\ee
Noticeable production of $e^+ e^-$ pairs by BH takes place if $E (r_g) \gtrsim E_c$, i.\,e. taking into account \eqref{condition_q_p} if
\be \label{condition_M}
M_{20} \lesssim \frac{q}{0.06} \lesssim 10. 
\ee
 
Since $E (r) = E (r_g) \times \left( r_g / r \right)^2$, the probability of pair production by a single BH as a function of distance $r$ is equal to
\be
W (r) = \left( \frac{2 q m_p M}{\pi m_{Pl}^2} \right)^2\, \frac{1}{r^4} \sum_\text{$n = 1$}^\infty \frac{1}{n^2} \exp \left( - \sqrt{\pi\,} n\, \frac{m_e^2 m_{Pl}^2}{4 q m_p M}\, r^2 \right).
\label{W-of-r}
\ee

Therefore, the expression for the number of $e^+ e^-$ pairs produced per unit time by BH with electric charge $q$ looks as follows:
\begin{align} \label{dotNee}
\dot N_{ee} & = \int d^3 r\, W (r) = \frac{1}{\pi} \left( \frac{4 q m_p M}{m_{Pl}^2} \right)^2 \int_{r_g}^\infty \frac{dr}{r^2} \sum_\text{$n = 1$}^\infty \frac{1}{n^2} \exp \left( - \sqrt{\pi\,} n\, \frac{m_e^2 m_{Pl}^2}{4 q m_p M}\, r^2 \right) = \nonumber \\
& = 8 m_e \left( \frac{q m_p M}{\sqrt{\pi\,} m_{Pl}^2} \right)^{3/2} \int_{z_g}^\infty \frac{dz}{z^2} \sum_\text{$n = 1$}^\infty \frac{\exp \left( - n z^2 \right)}{n^2},
\end{align}
where
\be
\label{z-of-q}
z = \frac{\sqrt[4]{\pi\,} m_e m_{Pl}}{2 \sqrt{q m_p M}}\, r \quad \text{and} \quad z_g = \frac{z}{r}\, r_g = \frac{\sqrt[4]{\pi\,} m_e}{m_{Pl}} \sqrt{\frac{M}{q m_p}} \approx 0.43\, \sqrt{\frac{M_{20}}{q}}.
\ee

The integral in Eq.~\eqref{dotNee} is dominated by $n = 1$ term and is equal to
\be
I(z_g) = \int_{z_g}^\infty \frac{dz}{z^2} \sum_\text{$n = 1$}^\infty \frac{\exp \left( - n z^2 \right)}{n^2} \approx \int_{z_g}^\infty \frac{dz}{z^2}\, e^{- z^2} = \frac{e^{- z_g^2}}{z_g} - \sqrt \pi\, \mathrm{erfc} \left( z \right),
\label{int}
\ee
where $\mathrm{erfc} \left( z \right) = (2/\sqrt \pi) \times \int_{z_g}^\infty e^{- z^2} dz$ is the complementary error function. 

Substituting the numerical values of constants into Eq.~\eqref{dotNee}, we obtain the following expression for the flux of positrons:
\be \label{dot-Ne-2}
\dot N_{ee} \approx \left( q M_{20} \right)^{3/2} I \left( 0.43 \sqrt{M_{20}/q}\, \right) \times 5.5 \times 10^{29} \text{ s}^{-1}.
\ee

\section{Results and discussion} \label{discussion}
 
In this Section we discuss whether primordial BHs with masses around $10^{20}$ g can produce enough positrons by the Schwinger mechanism to partially or completely explain the $e^+ e^-$ annihilation rate in the Galactic bulge \eqref{Neebulge}.

The dark matter (DM) mass in the Galactic bulge $M_\text{DM}^\text{bulge}$ amounts approximately from $0.17 \times 10^{10} M_\odot$ to $0.55 \times 10^{10} M_\odot$ depending on the model~\cite{bulge}, where $M_\odot \approx 2 \times 10^{33}$ g is the solar mass. Therefore, one can consider that
\be
M_\text{DM}^\text{bulge} \lesssim 10^{43} \text{ g},
\ee
and since the primordial BHs with masses close to $10^{20}$ g might make significant contribution, up to 100\,\%, to the DM~\cite{Carr-Kuhnel-1, Carr-Kuhnel-2}, the number of such BHs can be bounded as 
\be
N_\text{BH}^\text{bulge} = \frac{M_\text{DM}^\text{bulge}}{M} \lesssim \frac{10^{43} \text{ g}}{M_{20} \times 10^{20} \text{ g}} = \frac{10^{23}}{M_{20}}
\ee
and the density number as 
\be
n_\text{BH}^\text{bulge} = \frac{N_\text{BH}^\text{bulge}}{V^\text{bulge}} \lesssim \frac{10^{-43}}{M_{20}} \text{ cm}^{-3},
\ee
where we used that the volume of the Galactic bulge is $V^\text{bulge} \sim 30 \text{ kpc}^3 \sim 10^{66} \text{ cm}^3$~\cite{bulge}.

\begin{figure}[htb]
\centering
\includegraphics[scale = 0.7857]{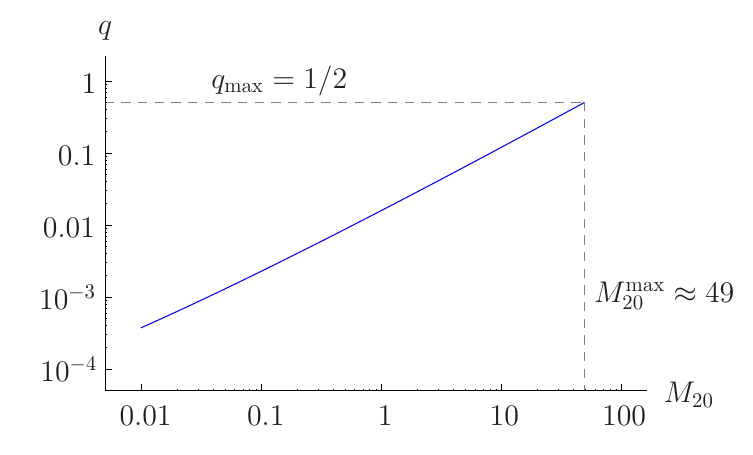} 
\caption{Dependence of BH charge on its mass in the case when all positrons in the Galactic bulge is produced by primordial BHs according to the Schwinger mechanism.}
\label{fig_qM20}
\end{figure}

Assume that a considerable part of the positrons in the Galactic bulge is produced by these BHs according to the Schwinger mechanism and all of them annihilate with electrons in cosmic plasma. Then taking into account \eqref{Neebulge} we come to the following estimate for the number of such positrons produced per unit time by a single charged BH:
\be \label{estNee}
\dot N_{ee} \sim \frac{\dot N_{ee}^\text{bulge}}{N_\text{BH}^\text{bulge}} \sim \frac{10^{43} \text{ s}^{-1}}{10^{23}/M_{20}} = M_{20} \times 10^{20} \text{ s}^{-1}.
\ee
Equating this expression to the relation \eqref{dot-Ne-2} and solving numerically the obtained equation,
\be \label{qM20}
\left( q M_{20} \right)^{3/2} I \left( 0.43 \sqrt{M_{20}/q}\, \right) \times 5.5 \times 10^9 = M_{20},
\ee
we get the corresponding dependence of BH charge on its mass, $q(M_{20})$, see Fig.~\ref{fig_qM20}. Note that according to \eqref{condition_final} there is a maximum BH mass $M_{20}^{\max} \approx 49$ corresponding to maximum BH charge $q_{\max} = 1/2$ at which protons do not fall into the BH due to the strong Coulomb repulsion. As for the lower bound on BH charge $q$, see \eqref{condition_final}, numerical calculations show that it always holds for the considered values of $M_{20} \gtrsim 0.01$.

\begin{figure}[htb]
\centering
\includegraphics[scale = 0.7857]{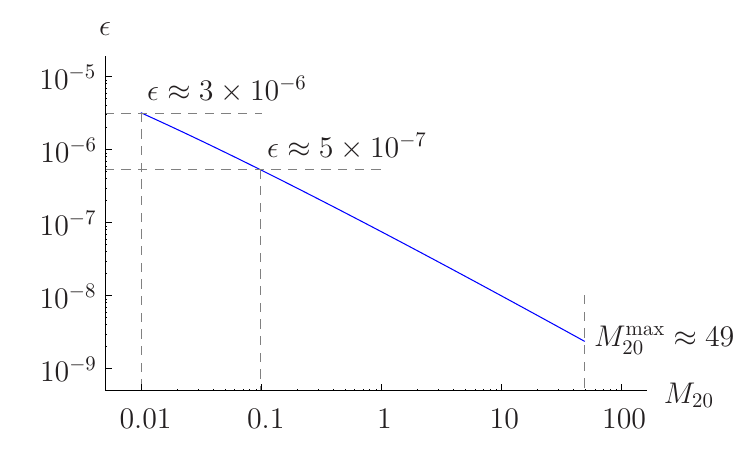} 
\caption{Dependence of proton-to-electron relative flux $\epsilon = (\dot N_p - \dot N_e)/\dot N_e$ on BH mass in the case when all positrons in the Galactic bulge is produced by primordial BHs according to the Schwinger mechanism.}
\label{fig_epsM20}
\end{figure}

Let us go back now to expression \eqref{dotNp} for the flux of protons captured by the BH. Taking into account the law of electric charge conservation and relation \eqref{k},
\be
\dot N_{ee} = \dot N_p - \dot N_e = (k - 1) \dot N_e,
\ee 
we obtain the following equation:
\be \label{eq_final}
\dot N_{ee} = \frac{k - 1}{k}\, \dot N_p,
\ee
or in other words,
\be
\frac{k - 1}{k} \times \frac{0.1}{\eta} \left( q(M_{20}) + 2.7 \times 10^{-4} \, \frac{k^2 - 1}{k^2} \right) \times 8.4 \times 10^8 = 1.
\ee
Assuming that radiative efficiency of accretion $\eta = 0.1$, see \cite{FKR}, and taking into account solution $q(M_{20})$ of Eq.~\eqref{qM20} we find that the proton-to-electron ratio $\dot N_p/\dot N_e = k$ is equal to 1 with very good accuracy, i.\,e. $\epsilon = k - 1 = (\dot N_p - \dot N_e)/\dot N_e \ll 1$. The dependence $\epsilon(M_{20})$ is shown in Fig.~\ref{fig_epsM20}.

\begin{figure}[htb]
\centering
\includegraphics[scale = 0.7857]{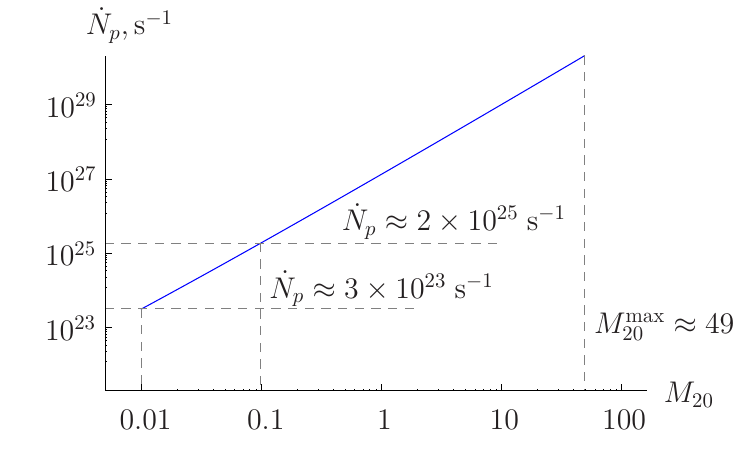} 
\caption{Dependence of proton flux on BH mass in the case when all positrons in the Galactic bulge is produced by primordial BHs according to the Schwinger mechanism.}
\label{fig_NpM20}
\end{figure}

Therefore, from relations \eqref{estNee} and \eqref{eq_final} it follows that fluxes of protons and cosmic plasma electrons captured by the BH are equal to
\be \label{NpNe}
\dot N_p \approx \dot N_e = \frac{\dot N_{ee}}{\epsilon} \sim \frac{M_{20}}{\epsilon} \times 10^{20} \text{ s}^{-1}.
\ee
The dependence $\dot N_p(M_{20})$ is shown in Fig.~\ref{fig_NpM20}.

Let us discuss whether such values of proton flow on BHs are possible. The process of the proton capture by BHs could start only after the galaxies were formed when matter decoupled from the overall Hubble expansion. Therefore, the ratio $\dot M/M$ can be bounded from above using the condition $\dot M \tau_\text{gal} \ll M$, where $\tau_\text{gal}$ is the time that passed from the galaxy formation. Since $\tau_\text{gal} \sim 10^{10} \text{ yr} \approx 3 \times 10^{17} \text{ s}$ we obtain that
\be \label{dotNp1}
\dot N_p \approx \frac{\dot M}{m_p} \lesssim \frac{M}{m_p \tau_\text{gal}} \sim 2 M_{20} \times 10^{26} \text{ s}^{-1}
\ee
and consequently that $\epsilon \gtrsim 5 \times 10^{- 7}$, see \eqref{NpNe}. From Fig.~\ref{fig_epsM20} one can see that it holds only for $M_{20} \lesssim 0.1$, i.\,e. $\dot N_p \lesssim 2 \times 10^{25} \text{ s}^{-1}$, see \eqref{dotNp1} and Fig.~\ref{fig_NpM20}. Therefore, $e^+ e^-$ production by primordial BHs with masses $\lesssim 0.1 \times 10^{20} \text{ g} = 10^{19}$~g can explain $e^+ e^-$ annihilation rate in the Galactic bulge \eqref{Neebulge} completely. As for BHs with masses $\gtrsim 10^{19}$ g, they can produce the number of positrons sufficient only to partially explain the rate \eqref{Neebulge}. However, we should take into account that Milky Way started to form from the low-density protocloud, thus initially the density of protons/electrons was much smaller and contemporary value of $\dot N_p$ may exceed the bound \eqref{dotNp1}. 

An upper limit on $\dot N_p$ can also be obtained from proton (electron) density in the Galaxy. About 90\,\% of the ionized gas in the Galaxy is low density ionized hydrogen in a warm ionised medium with an average electron density $n_e = 0.03$\,---\,$0.08 \text{ cm}^{-3}$ \cite{density}. In contrast H II regions with $n_e \gtrsim 100 \text{ cm}^{-3}$ are much less extensive \cite{density}. Assume that the average proton (electron) density in the Galactic bulge is $n_p^\text{bulge} \sim 1 \text{ cm}^{-3}$. Then there are 
\be
n_p^\text{bulge}/n_\text{BH}^\text{bulge} \sim M_{20} \times 10^{43}
\ee
times more protons in the bulge than BHs. Therefore, the rate of capture of protons by a BH should satisfy the condition
\be
\dot N_p \lesssim \frac{M_{20} \times {10^{43}}}{\tau_\text{gal}} \sim 3 M_{20} \times 10^{25} \text{ s}^{-1}.
\ee
From \eqref{NpNe} it follows that $\epsilon \gtrsim 3 \times 10^{- 6}$ and from Fig.~\ref{fig_NpM20} one can see that we get more stringent upper limit, $M_{20} \lesssim 0.01$, i.\,e. $M \lesssim 10^{18}$~g, and consequently that $\dot N_p \lesssim 3 \times 10^{23} \text{ s}^{-1}$, see \eqref{dotNp1} and Fig.~\ref{fig_NpM20}. Therefore, according to this estimate $e^+ e^-$ production by primordial BHs with masses in the considered range $\sim 10^{18}$\,---\,$10^{21}$ g can explain $e^+ e^-$ annihilation rate in the Galactic bulge \eqref{Neebulge} only partially. And the lighter BHs are, the greater contribution they can make.

A promising source of positrons could be the Galactic Center where the density of dark matter is much higher than the average galactic one. According to \cite{Sofue} it is as large as 840 GeV/cm$^3$. Moreover, the number density of stars in the Galactic Center is close to 10 million stars per cubic parsec. This density is by far larger than the number density of stars in the solar neighborhood, that is only 0.1\,---\,0.2 stars per cubic parsec. Hence a much higher than the average number density of protons can be naturally expected.

However, the case of heavier BHs, say, with $M_{BH} \sim 10^{22}$ g also looks promising. According to \cite{clump} and the subsequent one \cite{cluster} dark matter could form sub-galactic clumps where its density may be strongly enhanced, in particular, the density of BHs with the close to mentioned above values of masses. Such high density clusters may be the sources of the observed 511 keV photons probably originating from $e^+ e^-$ annihilation.

Seemingly the clusterization of primordial black holes (PBHs) is a general phenomenon, see e.g. \cite{ES}. Together with dark matter made of PBHs inside the clumps one may expect a higher than the galactic average density of the ordinary matter. So these clusters are potentially interesting sources of positrons.

There are many unknowns in the details of the Galaxy structure and depending on it the discussed here mechanism of the positron production could probably explain at least some part of 511 keV gamma quanta flux.

\section{Conclusions} \label{s-concl}

Primordial black holes (PBHs) with masses near $10^{20}$ g are attractive candidates for the cosmological dark matter. It would be interesting if these PBHs were simultaneously the sources of cosmic positrons as described in the present work. The considered here mechanism of positron production may explain some fraction of the observed 511 keV gamma radiation flow.

Of course one cannot exclude some other sources of low energy positrons. For example primordial antimatter or decays of light dark matter particles could be serious competitors. On the other hand, several independent sources of positrons are allowed. Each possible source of positrons has its specific features that allow to distinguish between them. Hopefully this can be achieved with more detailed future observations.

\section*{Funding}
This work was supported by the Russian Science Foundation under grant no.~22-12-00103.



\begin{thebibliography}{99}

\bibitem{Shv}
V.\,F. Shvartsman, 
\textit{Generation of relativistic particles by neutron stars in the state of accretion}, 
Astrophysics \textbf{6} (1970) 159, translated from Astrofizika \textbf{6} (1970) 309.

\bibitem{TZTI}
R. Turolla, S. Zane, A. Treves, and A. Illarionov, 
\textit{On electrostatic positron acceleration in the accretion flow onto neutron stars}, 
Astrophys. J. \textbf{482} (1997) 377, arXiv:astro-ph/9612001.

\bibitem{ZTT}
S. Zane, R. Turolla, and A. Treves, 
\textit{Hot atmospheres around accreting neutron stars: a possible source for hard X-ray emission}, 
Astrophys. J. \textbf{501} (1998) 258, arXiv:astro-ph/9706200.

\bibitem{TT}
A. Treves and R. Turolla, 
\textit{Vacuum breakdown near a black hole charged by hypercritical accretion}, 
Astrophys. J. \textbf{517} (1999) 396, arXiv:astro-ph/9812383.

\bibitem{BDP}
C. Bambi, A.\,D. Dolgov, and A.\,A. Petrov,
\textit{Black holes as antimatter factories}, 
JCAP \textbf{09} (2009) 013, arXiv:0806.3440 [astro-ph].

\bibitem{Schw}
J. Schwinger,
\textit{On gauge invariance and vacuum polarization},
Phys. Rev. \textbf{82} (1951) 664.

\bibitem{anti-e1}
J. Kn\"{o}dlseder \textit{et al.},
\textit{The all-sky distribution of 511 keV electron-positron annihilation emission},
Astron. Astrophys. \textbf{441} (2005) 513, arXiv:astro-ph/0506026.

\bibitem{anti-e2}
P. Jean \textit{et al.},
\textit{Spectral analysis of the Galactic $e^+ e^-$ annihilation emission},
Astron. Astrophys. \textbf{445} (2006) 579, arXiv:astro-ph/0509298.

\bibitem{anti-e3}
G. Weidenspointner \textit{et al.},
\textit{The sky distribution of positronium annihilation continuum emission measured with SPI/INTEGRAL},
Astron. Astrophys. \textbf{450} (2006) 1013, arXiv:astro-ph/0601673.

\bibitem{gr-ann}
I.\,F. Mirabel, 
\textit{The Great Annihilator in the Central Region of the Galaxy},
Messenger \textbf{70} (1992) 51.

\bibitem{Ein-observ}
P. Hertz and J.\,E. Grindlay,
\textit{The Einstein galactic plane survey: statistical analysis of the complete X-ray sample},
Astrophys. J. \textbf{278} (1984) 137.

\bibitem{granat}
R. Sunyaev \textit{et al.},
\textit{Two hard X-ray sources in 100 square degrees around the Galactic Center},
Astron. Astrophys. \textbf{247} (1991) L29.

\bibitem{AMS-19}
M. Aguilar \textit{et al.} (AMS Collaboration),
\textit{Towards Understanding the Origin of Cosmic-Ray Positrons},
Phys. Rev. Lett. \textbf{122} (2019) 041102.

\bibitem{DS}
A. Dolgov and J. Silk, 
\textit{Baryon isocurvature fluctuations at small scales and baryonic dark matter},
Phys. Rev. \textbf{D 47} (1993) 4244.

\bibitem{DKK}
A.\,D. Dolgov, M. Kawasaki, and N. Kevlishvili, 
\textit{Inhomogeneous baryogenesis, cosmic antimatter, and dark matter},
Nucl. Phys. \textbf{B 807} (2009) 229, arXiv:0806.2986 [hep-ph].

\bibitem{BBBDP}
A.\,E. Bondar, S.\,I. Blinnikov, A.\,M. Bykov, A.\,D. Dolgov, and K.\,A. Postnov,
\textit{X-ray signature of antistars in the Galaxy}, 
JCAP \textbf{03} (2022) 009, arXiv:2109.12699 [astro-ph.HE].

\bibitem{anti-nuc-AMS-1}
S. Ting, 
\textit{A Brief Summary of Ten Years of New and Unexpected Results from the Alpha Magnetic Spectrometer on the International Space Station}, 
proceedings of 44th COSPAR Scientific Assembly (16-24 July 2022), vol. 44, p. 3071.

\bibitem{anti-nuc-AMS-2}
V. Choutko, 
\textit{Cosmic Heavy Anti-Matter}, 
proceedings of 44th COSPAR Scientific Assembly (16-24 July 2022), vol. 44, p. 2083.

\bibitem{anti-stars}
S. Dupourqu{\'e}, L. Tibaldo, and P. von Ballmoos, 
\textit{Constraints on the antistar fraction in the Solar System neighborhood from the 10-year Fermi Large Area Telescope gamma-ray source catalog},
Phys. Rev. \textbf{D 103} (2021) 083016, arXiv:2103.10073 [astro-ph.HE].

\bibitem{bpbbd}
A. Bykov, K. Postnov, A. Bondar, S. Blinnikov, and A. Dolgov,
\textit{Antistars as possible sources of antihelium cosmic rays}, 
arXiv:2304.04623 [astro-ph.HE].

\bibitem{Laha-1}
R. Laha,
\textit{Primordial Black Holes as a Dark Matter Candidate Are Severely Constrained by the Galactic Center 511 keV $\gamma$-Ray Line},
Phys. Rev. Lett. \textbf{123} (2019) 251101, arXiv:1906.09994 [astro-ph.HE].

\bibitem{Laha-2}
B. Dasgupta, R. Laha, and A. Ray,
\textit{Neutrino and Positron Constraints on Spinning Primordial Black Hole Dark Matter},
Phys. Rev. Lett. \textbf{125} (2020) 101101, arXiv:1912.01014 [hep-ph].

\bibitem{Carr-Kuhnel-1}
B. Carr and F. K\"{u}hnel,
\textit{Primordial Black Holes as Dark Matter: Recent Developments},
Ann. Rev. Nucl. Part. Sci. \textbf{70} (2020) 355, arXiv:2006.02838 [astro-ph.CO].

\bibitem{Carr-Kuhnel-2}
B. Carr and F. K\"{u}hnel,
\textit{Primordial black holes as dark matter candidates},
SciPost Phys. Lect. Notes \textbf{48} (2022), arXiv:2110.02821 [astro-ph.CO].

\bibitem{Bondi}
H. Bondi,
\textit{On spherically symmetrical accretion},
Mon. Not. Roy. Astron. Soc. \textbf{112} (1952) 195.

\bibitem{FKR}
J. Frank, A. King, and D. Raine, 
\textit{Accretion Power in Astrophysics},
3rd Edition, Cambridge University Press, 2002.

\bibitem{bulge}
M. Portail, C. Wegg, O. Gerhard, and I. Martinez-Valpuesta,
\textit{Made-to-measure models of the Galactic box/peanut bulge: stellar and total mass in the bulge region},
Mon. Not. Roy. Astron. Soc. \textbf{448} (2015) 713, arXiv:1502.00633 [astro-ph.GA].

\bibitem{density}
W.\,D. Langer \textit{et al.},
\textit{The dense warm ionized medium in the inner Galaxy},
Astron. Astrophys. \textbf{651} (2021) A59, 	arXiv:2105.07023 [astro-ph.GA].

\bibitem{Sofue} 
Y. Sofue,
\textit{Rotation Curve of the Milky Way and the Dark Matter Density},
Galaxies \textbf{8} (2020) 37, arXiv:2004.11688 [astro-ph.GA].

\bibitem{clump}
P. Blasi, R. Dick, and E.\,W. Kolb, 
\textit{Ultra-high energy cosmic rays from annihilation of superheavy dark matter},
Astropart. Phys. \textbf{18} (2002) 57, arXiv:astro-ph/0105232.

\bibitem{cluster}
S.\,G. Rubin, A.\,S. Sakharov, and M.\,Yu. Khlopov, 
\textit{The Formation of Primary Galactic Nuclei during Phase Transitions in the Early Universe}, 
J. Exp. Theor. Phys. \textbf{92} (2001) 921, arXiv:hep-ph/0106187. 

\bibitem{ES}
Yu. Eroshenko and V. Stasenko, 
\textit{Gravitational Waves from the Merger of Two Primordial Black Hole Clusters},
Symmetry \textbf{15} (2023) 637, arXiv:2302.05167 [astro-ph.CO].

\end{thebibliography}
\end{document}